\documentclass[aps,prl,twocolumn,groupedaddress,longbibliography,nofootinbib]{revtex4-2}
\pdfoutput=1

\usepackage{lmodern} 
\usepackage{microtype} 
\usepackage[english]{babel}

\usepackage{color,graphicx}
\usepackage[usenames,dvipsnames]{xcolor}

\usepackage{changes}

\usepackage{amsmath,dsfont,url,amssymb}
\usepackage{slashed,cancel}
\usepackage[usenames,dvipsnames]{xcolor} 
\usepackage[colorlinks=true, linkcolor=black, citecolor=ForestGreen]{hyperref}
\usepackage{mdframed}
\usepackage{subfigure,overpic}

\DeclareMathOperator{\Tr}{Tr}
\renewcommand{\Im}[0]{\operatorname{Im}}
\renewcommand{\Re}[0]{\operatorname{Re}}

\begin{document}

\title{Damping of Pseudo-Goldstone Fields}

\author{Luca V.~Delacr\'etaz}
\email{lvd@uchicago.edu}
\affiliation{Kadanoff Center for Theoretical Physics, University of Chicago, Chicago, IL 60637, USA}
\author{Blaise Gout\'eraux}
\email{blaise.gouteraux@polytechnique.edu}
\affiliation{CPHT, CNRS, École Polytechnique, IP Paris, F-91128 Palaiseau, France}
\author{Vaios Ziogas}
\email{vaios.ziogas@polytechnique.edu}
\affiliation{CPHT, CNRS, École Polytechnique, IP Paris, F-91128 Palaiseau, France}

\date{\today}

\begin{abstract}
Approximate symmetries abound in Nature. If these symmetries are also spontaneously broken, the would-be Goldstone modes acquire a small mass, or inverse correlation length, and are referred to as pseudo-Goldstones. At nonzero temperature, the effects of dissipation can be captured by hydrodynamics at sufficiently long scales compared to the local equilibrium. Here we show that in the limit of weak explicit breaking, locality of hydrodynamics implies that the damping of pseudo-Goldstones is completely determined by their mass and diffusive transport coefficients. We present many applications: superfluids, QCD in the chiral limit, Wigner crystal and density wave phases in the presence of an external magnetic field or not, nematic phases and (anti-)ferromagnets. For electronic density wave phases, pseudo-Goldstone damping generates a contribution to the resistivity independent of the strength of disorder, which can have a linear temperature dependence provided the associated diffusivity saturates a bound. This is reminiscent of the phenomenology of strange metal high $T_c$ superconductors, where charge density waves are observed across the phase diagram.
\end{abstract}
\maketitle


 In the simplest setting, collective excitations associated with Goldstone modes for spontaneously broken symmetries disperse linearly $\omega(q)\simeq \pm c_s q$. At finite temperature, dissipation enters to leading order in the wavevector $q$ as a diffusive broadening of this ballistic sound mode $\omega(q) \simeq \pm c_s q - i D q^2/2$. The softness of these modes as $q\to 0$ is protected by the symmetry, and its spontaneous breaking. When the symmetry is only approximate, the collective excitation is sometimes referred to as a pseudo-Goldstone mode, and acquires a finite correlation length $1/q_o$. At finite temperature, it can also have a non-zero relaxation rate, so that the pole is located schematically at $\lim_{q\to 0}\omega(q) \simeq\pm c_s q_o - i\Omega/2$. The main result of this paper is to show that these transport coefficients are not all independent in the limit of weak symmetry breaking $q_o\to 0$. We derive the following relation for the pseudo-Goldstone damping rate
\begin{equation}\label{eq_relation}
\Omega = q_o^2 D + O(q_o^4)\, .
\end{equation}
Strictly speaking, there are typically several contributions to the attenuation $D$ and to the damping $\Omega$, which can be obtained using various Kubo formulas; we will clarify 
below
which ones satisfy a relation of the form \eqref{eq_relation}.

In the hydrodynamic regime of thermalizing systems and away from thermal phase transitions, there are no long-range excitations other than those accounted for by global symmetries. A basic property that follows is locality of constitutive relations, i.e.~expansions of operators (typically currents) in terms of conserved densities or Goldstone fields, and sources. As we will show, when the thermal correlation length is large (which happens when an approximate symmetry is spontaneously broken) the condition that sources enter locally in constitutive relations is not automatically satisfied in the Kadanoff-Martin approach to computing hydrodynamic response functions \cite{kadanoff1963hydrodynamic} and must be imposed by hand, leading to \eqref{eq_relation}. The breakdown of locality is also crisply encapsulated in the non-commutativity between the $q\to0$ and $q_o\to0$ limits in the Kubo formulas that define $\Omega$ and $D$. Restoring locality with \eqref{eq_relation} restores the commutativity of limits.

Given the prevalence of Goldstones for approximate symmetries in nature and in experiments, relations of the form \eqref{eq_relation} have many applications; we survey a variety of them: superfluids, QCD in the chiral limit, Wigner crystals including in the presence of an external magnetic field, nematic phases and (anti-)ferromagnets; in so doing, we extend our result to Goldstone excitations with more complicated dispersion relations. We comment on the implications of \eqref{eq_relation} for strange metallic transport in high temperature superconductors (HTSC). In the main text, we give a detailed presentation for the case of a global $U(1)$ symmetry and only sketch other cases. We provide many technical details in a set of appendices.

Damping of pseudo-Goldstones at finite temperature has been studied for some time \cite{PhysRevB.17.535,RevModPhys.60.1129,PhysRevB.62.7553,PhysRevB.18.6245,Son:2001ff,Son:2002ci,Delacretaz:2016ivq,Delacretaz:2017zxd,Jokela:2017ltu,Andrade:2017cnc,Alberte:2017cch, Delacretaz:2019wzh}, although the relation \eqref{eq_relation} was not recognized at the time. Instead, to the best of our knowledge, \eqref{eq_relation} was first demonstrated in a Gauge/Gravity duality model of translation symmetry breaking, \cite{Amoretti:2018tzw}, see also for subsequent studies \cite{Donos:2019txg, Andrade:2018gqk,Ammon:2019wci,Baggioli:2019abx,Donos:2019hpp,Amoretti:2019kuf,Baggioli:2020edn,Andrade:2020hpu,Donos:2021ueh,Amoretti:2021fch,Amoretti:2021lll,Donos:2021pkk,Ammon:2021slb}. \eqref{eq_relation} was also noted in the hydrodynamic description of soft pions in \cite{Grossi:2020ezz}. Previous efforts towards proving \eqref{eq_relation} include \cite{Baggioli:2020nay,Baggioli:2020haa,Amoretti:2021fch}, and we comment on them in the relevant appendices.

In the absence of explicit breaking, Goldstones can be damped by the proliferation of vortices (and more generally topological defects for the spontaneously broken symmetry) which relax the winding of the phase \cite{PhysRev.140.A1197,PhysRevLett.41.121} and contribute to $\Omega$, \cite{Davison:2016hno,Delacretaz:2017zxd}. Here we will assume that vortices do not play any significant role and in particular that we are far away from any melting transition, so that $\Omega$ is perturbatively activated when the symmetry is weakly explicitly broken.

\section{Locality of constitutive relations}\label{ssec_locality}

In thermalizing systems, the slow excitations which govern the late time dynamics are long-wavelength modulations of conserved densities, satisfying continuity relations of the form 
\begin{equation}\label{eq_continuity}
\dot n_a(x,t) + \nabla \cdot j_a(x,t) = 0\, , 
\end{equation}
where $a=1,2,\ldots$ labels the various densities, and dot denotes a time derivative. The current densities $j_a$ are not themselves slow operators -- their time derivatives are not suppressed by gradients -- but as any operator they can be effectively expanded locally in terms of the slow densities at late times. 
These expansions are referred to as constitutive relations; to linear order in the densities they read schematically%
	\footnote{We shall ignore nonlinear hydrodynamic fluctuations in this work. }
\begin{equation}\label{eq_constitutive}
j = \alpha_0 n + \alpha_1 \nabla n +  \alpha_2 \nabla^2 n + \cdots\, , 
\end{equation}
where $\alpha_i$ are a priori unknown coefficients. This expression should be understood as an effective operator equation: although it is not valid microscopically, correlation functions involving operators of either side of the equation match in the hydrodynamic regime. The conservation law \eqref{eq_continuity} then produces an equation of motion for the densities
\begin{equation}\label{eq_eom0}
\dot n_a(q,t) + M_{ab}(q)n_b(q,t) = 0\, , 
\end{equation}
where we have defined $n(q,t) = \int d^d x \, e^{iqx} n(x,t)$, and summation over repeated indices is implied. As usual, external sources are introduced by deforming the Hamiltonian
\begin{equation}\label{eq_H}
H_0\to H(t) = H_0 - \int d^d x \, \delta \mu^a(x,t) n_a(x,t)\, ,
\end{equation}
and enter in the conservation equations as (see appendix):
\begin{equation}\label{eq_eom}
\dot n_a(q,t) + M_{ab}(q)\bigl(n_b(q,t) - \chi_{bc}(q)\delta \mu_c(q,t)\bigr) = 0 \, .
\end{equation}
$\chi(q)$ is the matrix of static susceptibilities,
\begin{equation}\label{eq_susceptibility}
\chi_{ab}(x-x')
	\equiv - \frac{\delta^2 W}{\delta \mu^a(x) \delta \mu^b(x')}\, ,
\end{equation}
where $ W =- T \log Z = -T \log \Tr e^{-\beta H}$ is the equilibrium thermal free energy, and $H$ is given by \eqref{eq_H} but with time independent sources $\delta\mu(x,t)=\delta\mu(x)$. The matrices $\chi$ and $M$ satisfy various positivity and symmetry conditions due to, e.g., positivity of dissipation or Onsager relations -- these are well known and we will not comment further upon them.

We are now ready to state our constraint: Eq.~\eqref{eq_eom} must be a (sufficiently) local function of the densities and sources. By local, we mean that it must satisfy an expansion in momentum $q$, with higher powers of $q$ suppressed by a scale which sets the cutoff of hydrodynamics. The matrix $M_{ab}(q)$ satisfies this condition, since it originated from the local constitutive relation \eqref{eq_constitutive}. What about $M_{ab}(q)\chi_{bc}(q)$? Static susceptibilities are local up to the thermal correlation length $\xi$ (sometimes called inverse thermal mass). This length scale is usually smaller than the length scales of interest in hydrodynamics, except in two situations: close to thermal phase transitions, and in the presence of Goldstone or pseudo-Goldstone modes.\footnote{One more exotic situation where this may arise is in systems with approximate dipole conservation \cite{Grosvenor:2021rrt,Glorioso:2021bif}.} In these situations, locality of \eqref{eq_eom} in the hydrodynamic regime is not automatic, and can lead to constraints on transport parameters. We derive such a constraint in the concrete example of a conserved $U(1)$ in the following section.

Locality of constitutive relations reflects the central assumption of hydrodynamics in thermalizing systems, namely that all long-lived and long-range excitations are accounted for by symmetries (as densities or Goldstone modes), and these are solely responsible for non-analyticities in thermal response functions at small wavevector $q$ and frequency $\omega$. Since none of these modes have been integrated out when writing expressions such as \eqref{eq_constitutive} (or \eqref{eq_eom} with sources), these expressions must be local.

\section{Implication for pseudo-Goldstone modes}\label{ssec_minimalex}
The simplest setting involving a pseudo-Goldstone field is a system with a single approximate continuous symmetry ($U(1)$ or $\mathbb R$) which is spontaneously broken. This could describe, e.g., the ordered phase of the XY model in the presence of a small symmetry breaking deformation. At finite temperature the system will thermalize and hydrodynamics will emerge -- the hydrodynamic modes are the charge density $n$ associated with the symmetry, its conjugate the Goldstone phase $\phi$, energy and momentum densities. Energy and momentum densities will not play an important role here and we shall ignore them (or assume they decouple) for simplicity.

The hydrodynamics of this system fits in the general framework established in the previous section, with $n$ and $\phi$ playing the role of densities.%
	\footnote{$\nabla\phi$ is in fact the density of a higher-form symmetry associated with conservation of winding, with the Josephson relation corresponding to the constitutive relation for the higher-form current $\dot \phi$ \cite{Grozdanov:2018ewh,Delacretaz:2019brr}.} 
Consider first the situation where the symmetry is exact, and the hydrodynamic theory is well known \cite{chaikin1995principles}. The matrix of susceptibilities \eqref{eq_susceptibility} is obtained by coupling the theory to static sources $H_0 \to H$,
\begin{equation}
 H= H_0 - \int d^d x \, \delta \mu(x) n(x,t) + \delta s_\phi(x) \phi(x,t)\, .  
\end{equation}
Since the thermal correlation length diverges due to the presence of the Goldstone, it is convenient not to integrate out the Goldstone phase $\phi$ \cite{Bhattacharyya:2012xi}, so that instead of working with $W[\delta\mu,\delta s_\phi]$ we will consider the free energy $F$ defined as
\begin{equation}\label{eq_WtoF}
e^{-\beta W}
	= \int D\phi \, e^{-\beta F[\delta\mu,\delta s_\phi,\phi]}\, .
\end{equation}
Only gradients of $\phi$ can appear in $F$, since $\phi$ shifts under the symmetry.
Let us now assume the symmetry is weakly broken. This will introduce a new length scale $1/q_o$ in the system, the thermal correlation length of $\phi$, which is parametrically larger than the cutoff length of hydrodynamics -- a hydrodynamic description of the system should exist that is valid across this new scale. A lower-gradient, symmetry-breaking term is allowed in $F = \int d^d x f$ :
\begin{equation}\label{eq_superfluid_F_broken}
f = \frac{f_s}2 [(\nabla \phi)^2 + q_o^2 \phi^2] - \delta s_\phi \phi  - \frac{\chi_{nn}}2 \delta\mu^2  + \cdots\, ,
\end{equation}
with $\chi_{nn}$ the charge susceptibility and $f_s$ the superfluid stiffness.
By integrating out $\phi$ and using \eqref{eq_WtoF}, we obtain the susceptibility matrix
\begin{equation}\label{eq_chi_superfluid}
\chi(q) \simeq  \left(\begin{array}{cc}
\chi_{nn}&0\\
0&\frac{1}{f_s (q^2+q_o^2)}\\
\end{array}\right)\, .
\end{equation}
Let us also review how the constitutive relations and conservation laws change \cite{Davison:2016hno,Delacretaz:2017zxd}, see also \cite{Ammon:2021slb}. To leading order in gradients, the most general way the conservation law can be weakly broken is
\begin{equation}
\label{U1currentbrokenconseq}
\dot n + \nabla \cdot j = - \Gamma n + f_s q_o^2 \phi + \cdots\, , 
\end{equation}
where we introduced a charge relaxation rate $\Gamma$. The last term is fixed by the symmetry breaking term in the free energy \eqref{eq_superfluid_F_broken}.\footnote{We fix the sign of the Goldstone field with the convention $[\int d^d x \, n(x),\phi(x')] = i$.} In the absence of sources, the constitutive relation for the current and the Josephson relation are\footnote{A different choice of hydrodynamic frame would lead to analogous constraints; see the appendix for further details.}
\begin{equation}
  j\simeq f_s \nabla \phi - D_n \nabla n\,,\quad\dot \phi \simeq -\Omega \phi- \frac{1}{\chi_{nn}}n + D_\phi \nabla^2 \phi  \, .
\end{equation}
In the absence of explicit symmetry breaking, the coefficients of the leading terms are fixed in terms of the coefficients appearing in the free energy \eqref{eq_superfluid_F_broken} with $q_o=0$, while two transport coefficients $D_n$, $D_\phi$ appear at subleading order in gradients. The $M$ matrix defined by \eqref{eq_eom} is given by
\begin{equation}\label{eq_Mmatrix_superfluid}
M(q)\simeq  \left(\begin{array}{cc}
\Gamma + D_n q^2 &-f_s(q_o^2 + q^2)\\
\frac{1}{\chi_{nn}}&\Omega + D_{\phi}q^2\\
\end{array}\right)\, .
\end{equation}
and is local. However, the other matrix appearing in the hydrodynamic equation of motion in the presence of sources \eqref{eq_eom}
\begin{equation}
M(q)\cdot\chi(q)
	\simeq \left(\begin{array}{cc}
\chi_{nn}(\Gamma + D_n q^2) &-1\\
1&\frac{\Omega +  D_{\phi}q^2}{f_s(q_o^2 + q^2)}\\
\end{array}\right)\,,
\end{equation}
is generically not -- the last term has an expansion as $q\to 0$ with higher powers of $q$ suppressed by $q_o$ instead of the hydrodynamic cutoff, which is parametrically larger. Locality is only restored if the transport parameters satisfy the relation 
\begin{equation}
\label{OmegaqoU1}
\Omega \simeq q_o^2 D_{\phi}\, , 
\end{equation}
to leading order in $q_o$. This relation was recently checked in two holographic models, \cite{Donos:2021pkk,Ammon:2021slb}.

Restoring locality through \eqref{OmegaqoU1} implies that the order of limits $q\to0$ and $q_o\to0$ commutes in the Kubo formulas defining $D_\phi$ and $\Omega$:
\begin{equation}
\label{Kubo_Dphi}
    D_\phi=f_s\lim_{\omega\to0}\lim_{q\to0}\lim_{\Gamma\to0}\lim_{q_o\to0}\frac1\omega\textrm{Im}G^R_{\partial_t\phi\partial_t\phi}(\omega,q)
\end{equation}
\begin{equation}
\label{Kubo_Omega}
    \frac{\Omega}{q_o^2}=f_s\lim_{\omega\to0}\lim_{\Gamma\to0}\lim_{q_o\to0}\lim_{q\to0}\frac1\omega\textrm{Im}G^R_{\partial_t\phi\partial_t\phi}(\omega,q)
\end{equation}
The limits $\Gamma\to0$ and $q_o\to0$ can be taken in any order.

In the appendix, we re-derive \eqref{OmegaqoU1} using the Schwinger-Keldysh formalism for effective theories of hydrodynamics \cite{kamenev2011field,Crossley:2015evo, Haehl:2015pja,Jensen:2018hse}. This is advantageous as locality is built-in from the start when constructing the effective action. There are only two independent symmetry breaking terms in the action, with coefficients $\omega_o$ and $\Gamma$, and \eqref{OmegaqoU1} follows automatically.

With \eqref{OmegaqoU1} in hand and turning external sources $\delta \mu,\, s_\phi$ back on, we observe that the Josephson relation can be rewritten as
\begin{equation}
\dot \phi \simeq\delta\mu- \frac{n}{\chi_{nn}} + \frac{D_\phi}{f_s} \left(s_\phi-h_\phi\right)\, ,
\end{equation}
where $ h_\phi\equiv\delta f/\delta\phi=f_s q_o^2\phi-f_s\nabla^2\phi$ is the field conjugate to $\phi$. In static equilibrium, $\langle h_\phi\rangle=s_\phi$. In practice, writing out the dissipative terms in constitutive relations in terms of the conjugate fields directly leads to local equations of motion. In [\citealp{Jensen:2012jh},\,\citealp{Bhattacharyya:2012xi}], it was emphasized how symmetry and consistency with static equilibrium with external sources constrains constitutive relations.


\section{QCD}\label{ssec_pions}

At temperatures below the chiral phase transition, the hydrodynamics of QCD includes pions as long-lived degrees of freedom, see e.g.~[\citealp{Son:2001ff},\,\citealp{Son:2002ci},\,\citealp{Jain:2016rlz},\,\citealp{Grossi:2020ezz}]. The spontaneously broken $SU(2)$ symmetry is only approximate, due to the quark masses; this case thus falls into the class of systems considered in this paper. At linear order in fields, the Josephson relation for pions is identical to the abelian one studied 
above. Imposing that the hydrodynamic equations of motion be local in the presence of sources, one thus finds the relation \eqref{OmegaqoU1} between the pion thermal mass, diffusivity, and relaxation rate.

This relation was in fact noticed recently in the context of QCD in \cite{Grossi:2020ezz}, where it was shown to follow from positivity of entropy production. This argument however does not straightforwardly apply to Goldstones for abelian symmetries, because additional terms can be added to the entropy current to guarantee positivity of entropy production without imposing the relation \eqref{OmegaqoU1}.\footnote{However, as shown in \cite{Armas:2021vku}, coupling the fields to external sources fixes this ambiguity and the entropy production argument applies.} Instead, the locality argument presented here applies to all of these situations.

\section{Goldstones for translation}\label{ssec_cdw}

The free energy density for an isotropic Wigner crystal in mechanical equilibrium in two spatial dimensions is \cite{chaikin1995principles}:
\begin{equation}
\label{freenergywc}
f=\frac{B-G}2 \left(\nabla^lu_l\right)^2+G\left(\nabla_{(i}u_{j)}\right)^2+\frac12G q_o^2 u^2\,,
\end{equation}
where the $u_i$ are the displacements, $B$ and $G$ the bulk and shear elastic moduli, and indices run over the spatial dimensions. The term proportional to $q_o^2$ is assumed to be small and explicitly breaks translation symmetry.

Momentum is relaxed:
\begin{equation}
    \dot\pi^i+\nabla_j\tau^{ji}=-\Gamma\pi^i-G q_o^2 u^i\,,
\end{equation}
while the Josephson equation is\footnote{We neglect charge and heat fluctuations. We give a more complete analysis in the appendix. Both the electric and heat currents receive new dissipative terms when $q_o\neq0$.}
\begin{equation}
\label{joseqwc}
 \dot u_i =\frac{\pi_i}{\chi_{\pi\pi}} -\Omega_{ij}u^j +D_\parallel \nabla_{i} \nabla^j u_j +D_\perp \epsilon_{ij}\nabla^{j} \nabla\times u\,.
\end{equation}
Locality of the $M\cdot\chi$ matrix following from \eqref{freenergywc}-\eqref{joseqwc} constrains both the damping and the diffusivities (see appendix)
\begin{equation}
\label{dampingWC}
    \frac{D_\parallel}{B+G}=\frac{D_\perp}{G}\equiv\frac{D}G\,,\quad \Omega_{ij}= q_o^2 D\,\delta_{ij}\,,
\end{equation}
as reported in holographic models of broken translations, \cite{Amoretti:2018tzw,Ammon:2019wci,Donos:2019hpp}.

\section{Strange metallic transport}

There is mounting experimental evidence that dynamical charge fluctuations play an important role across the phase diagram of cuprate HTSC \cite{Arpaia:2021}, which could play an important role in strange metallic transport, \cite{Grilli:2017,Delacretaz:2016ivq,Amoretti:2018tzw}.

Restoring charge and heat fluctuations and assuming approximate invariance under Galilean boosts, the resistivity is (see appendix)
\begin{equation}
\label{resistivityWCrelaxed}
    \rho_{\rm dc}=\frac{m^\star}{ne^2}\left(\Gamma+\frac{\omega_o^2}{\Omega}\right)=\frac{m^\star}{ne^2}\left(\Gamma+\frac{c_s^2}{ D}\right)\,,
\end{equation}
with $n$ the density, $e$ the unit charge, $m^\star$ the effective mass, $c_s=\sqrt{G/(m^\star n)}$ the Goldstone sound velocity, and $D$ defined as in \eqref{dampingWC}. 
\eqref{dampingWC} produces a finite resistivity, which is rooted in the relaxation of the pseudo-Goldstone in the bath of thermal excitations. We expect this result to extend to weakly-pinned electronic charge density wave systems where the Fermi surface is fully gapped and some electrons remain uncondensed, providing a bath of gapless excitations and giving rise to \eqref{dampingWC}. By contrast, if the Fermi surface is fully gapped, the damping \eqref{dampingWC} will be exponentially suppressed, recovering earlier hydrodynamic descriptions of pinned charge density waves \cite{RevModPhys.60.1129}.

This result straightforwardly extends to a unidirectional charge density wave. Since $\Gamma$ is typically $O(g^2)$ in the strength $g\ll1$ of explicit breaking, while $c_s^2/D$ is $O(g^0)$, the resistivity is large in the `hard' direction of the spontaneous modulation, small in the transverse `easy' direction, in general agreement with transport experiments in 2DES \cite{PhysRevLett.82.394} and in HTSC \cite{PhysRevLett.88.137005}. 

Combining \eqref{resistivityWCrelaxed} with arguments \cite{Hartnoll:2014lpa} that diffusivities are bounded from below by the Planckian timescale \cite{zaanen_superconductivity:_2004} in strongly-correlated materials
\begin{equation}
    D \simeq\frac{ \hbar}{k_B T}v^2\simeq\frac{ \hbar}{ k_B T}c_s^2\, ,
\end{equation}
(where for simplicity we took the characteristic velocity  $v=c_s$) leads to
\begin{equation}
    \rho_{\rm dc}\simeq\frac{m^\star}{ne^2}\left(\Gamma+\frac{k_B T}{\hbar}\right)\, .
\end{equation}
This gives a natural mechanism for the ubiquitous $T$-linear resistivity in these materials. Moreover, the slope of the linear term is independent from the strength of explicit breaking, as observed in experiments where samples are disordered by ion irradiation \cite{PhysRevLett.91.047001}. In contrast, $\Gamma$ originates from more conventional scattering mechanisms (Umklapp, disorder), which allows to account for experimental reports of competing scattering mechanisms with distinct temperature dependencies, \cite{cooper2009anomalous, Hussey:2011}.

\section{Wigner crystals in a magnetic field}

When placed in a magnetic field, the longitudinal and transverse hydrodynamic modes of a 2d Wigner crystal couple. The magnetophonon sector is described by the effective Lagrangian [\citealp{Watanabe:2013iia}, \citealp{Delacretaz:2019wzh}] 
\begin{equation}\label{eq_L}
\mathcal L
	\simeq \frac12 \omega_c \epsilon^{ij} {\varphi_i}\dot\varphi_j - \frac{1}{2} \varphi_i \left[\omega_o^2\delta^{ij} - \mu^{ijab}\nabla_a\nabla_b \right]\varphi_j\, .
\end{equation}
The presence of a term with a single time derivative reflects the breaking of time reversal due to a background magnetic field and leads to the canonical commutation relation between the Goldstones
\begin{equation}\label{eq_com_2}
[\varphi_i(x),\varphi_j(y)] = -\frac{i\epsilon_{ij}}{\omega_c} \delta(x-y)
\, ,
\end{equation}
which are no longer independent degrees of freedom. 

In the absence of pinning ($\omega_o=0$) $\varphi$ has to appear with a gradient in the potential because of invariance under translations $\varphi^i\to \varphi^i+c^i$. The most general stiffness tensor consistent with isotropy and PT symmetry is $\mu_{ijab}q^a q^b =B q_i q_j + G q^2\delta_{ij}$,
where the stiffnesses must satisfy $B,\,G>0$ for the potential to be positive definite. The $\varphi_i$ static susceptibility matrix takes the same form as in the case without a magnetic field. 

Locality enforces the following dissipative Josephson relation:
\begin{equation}
\label{josephsonmagnetophonon}
\dot \varphi_i \simeq-\left(\frac{\epsilon_{ij}}{\omega_c}+D_\varphi\delta_{ij}\right) \left( \delta_{jk}\omega_o^2- \mu^{jkab}\nabla_a\nabla_b  \right)  \varphi_k\, .
\end{equation}
Both the longitudinal  $\Omega=D_\varphi \omega_o^2$ and Hall $\Omega^H=\omega_o^2/\omega_c$ relaxation rates obey a relation analogous to \eqref{eq_relation} and are completely determined by $\omega_o$ and parameters of the unpinned theory. Where applicable, our results are consistent with  \cite{Amoretti:2021fch,Amoretti:2021lll,Donos:2021ueh}. The type II nature of the Goldstones is manifest in their dispersion relation ($q_y=0$) without pinning
\begin{equation}
    \omega\simeq\pm q_x^2\sqrt{\frac{G(B+G)}{\omega_c^2}-\frac{B^2D_\varphi^2}{4}}-\frac{i}{2}(B+2G)D_\varphi q_x^2\, ,
\end{equation}
or with pinning
\begin{equation}
\omega\simeq\pm\frac{\omega_o^2}{\omega_c} -iD_\varphi \omega_o^2\pm \frac{B+2G}{2\omega_c}q_x^2-\frac{i}{2}(B+2G)D_\varphi q_x^2\, .
\end{equation}

\section{Nematic phases}

Consider a phase in a translation invariant isotropic system where rotations are spontaneously broken. Examples include nematic or hexatic liquid crystals \cite{PhysRevB.19.2457,PhysRevB.22.2514} and quantum Hall systems \cite{PhysRevB.59.8065} (spontaneous breaking of a discrete rotation symmetry also arises in the cuprates \cite{Kivelson1998}). Spontaneous breaking of isotropy produces a Goldstone field that shifts under rotations $\theta\to \theta + c$. Since the generator of rotations is given by $J = \int d^2 x \, \epsilon_{ij}x^i \pi^j$, where $T_{0i}=\pi_i$ is the momentum density, this realization of the symmetry is implemented by the commutation relation
\begin{equation}\label{eq_theta_com}
[\pi_i(x), \theta(y)] = \frac12 i\epsilon_{ij} \partial_j \delta^2(x-y) + \cdots\, , 
\end{equation}
where $\cdots$ denote terms that are linear in $\theta$.  The hydrodynamic equations are well-known \cite{PhysRevB.22.2514,chaikin1995principles}. When rotation symmetry is only approximate, $\theta$ acquires a small gap $q_o^2$ as in the examples in previous sections, which can be measured in the susceptibility
\begin{equation}
\lim_{\omega\to 0} G^R_{\theta\theta}(\omega,q)
	= \chi_{\theta\theta}(q) \simeq \frac{1}{f_\theta}\frac{1}{q^2 + q_o^2}\, ,
\end{equation}
where $f_\theta$ is the stiffness which appear in the free energy as in \eqref{eq_superfluid_F_broken}.
The Josephson relation and stress tensor constitutive relation are
\begin{subequations}
\begin{align}
\dot \theta  \label{eq_thetaJo}
	&\simeq  - q_o^2 D_\theta  \theta +\frac12 \frac{1}{\chi_{\pi\pi}} \nabla \times \pi + D_\theta \nabla^2 \theta  \, , \\
\tau_{ij} \label{eq_thetaTij}
	&\simeq P    \delta_{ij} + \frac{f_{\theta}}{2}\epsilon^{ij}(\nabla^2-q_o^2) \theta +\pi_{ij}
\end{align}
\end{subequations}
with $\pi_{ij}=-D_\eta(\partial_i \pi_j + \partial_j \pi_i - \delta_{ij} \nabla\cdot \pi)- D_\zeta\nabla\cdot \pi \delta_{ij}$. 
The damping term $\Omega= q_o^2 D_\theta$ in \eqref{eq_thetaJo} follows from locality and obeys a relation analogous to \eqref{eq_relation}. The $q_o^2$ term in \eqref{eq_thetaTij} follows from consistency with \eqref{eq_theta_com}. Differently from previous examples, there is no analog of a relaxation rate $\Gamma$ for the angular momentum density: assuming that only rotation symmetry is broken but translation symmetry is preserved, the stress tensor is still conserved. One could of course lift this assumption and also break translations. Since $\Omega = q_o^2 D_\theta$, there is only one parameter introduced by the weak breaking of isotropy: $q_o$.

At the linear level, the Goldstone only mixes with the transverse momentum density $\nabla\times \pi$ and the longitudinal sector is unchanged from regular hydrodynamics; we will therefore focus on the transverse sector $\{\pi_\perp,\,\theta\}$. Solving the continuity relations, one finds when $q_o=0$ that this pair of modes disperses as
\begin{equation}
\omega_\pm \simeq - \frac{i}2 q^2 \left(D_\eta + D_\theta \pm \sqrt{(D_\eta-D_\theta)^2 - \frac{f_\theta}{\chi_{\pi\pi}}}\right).
\end{equation}
When $(D_\eta-D_\theta)^2 \geq {f_\theta}/{\chi_{\pi\pi}}$, these form two diffusive modes $\omega \sim -iq^2$, while in the opposite case $\omega_{\pm} \sim \pm q^2 - i q^2$.

When $q_o\neq0$, transverse momentum diffuses
\begin{equation}
   \omega_-\simeq-iq^2\left(D_\eta+\frac{f_\theta}{4 \chi_{\pi\pi}D_\theta}\right)\, ,
\end{equation}
while the Goldstone relaxes
\begin{equation}
    \omega_+\simeq-i D_\theta q_o^2-iq^2\left(D_\theta-\frac{f_\theta}{4 \chi_{\pi\pi}D_\theta}\right)\, .
\end{equation}

\section{(Anti-)ferromagnets}

Ferromagnets spontaneously break $SU(2)$ spin symmetry -- which can be viewed as an internal symmetry in the non-relativistic limit\footnote{See \cite{Gallegos:2021bzp} for a recent formulation of relativistic spin hydrodynamics} -- down to $U(1)$, with a finite magnetization $\langle n_a\rangle = M_0 \delta_a^3$. Antiferromagnets have the same symmetry breaking pattern, but $M_0=0$; their hydrodynamics is structurally similar to that of pions discussed above, and the results there apply with minor modifications. In this section we therefore focus on ferromagnets and briefly mention the constraints obtained from locality, leaving a more detailed study for future work. 

In the absence of explicit breaking of $SU(2)$ symmetry, magnons disperse as \cite{PhysRev.188.898}
\begin{equation}
\omega = \pm \frac{f_s}{M_0} q ^2 - i \gamma q^4 + \cdots\, .
\end{equation}
In practice, the $SU(2)$ symmetry is always approximate, and broken by spin-orbit effects. This leads to several possible explicit symmetry breaking scenarios: for example there may or may not be an unbroken $U(1)$, which may or may not be aligned with the magnetization. For the purposes of illustration, we focus on the situation where the entire $SU(2)$ is weakly explicitly broken. This will generate a finite magnon correlation length $1/q_o$, allow for a relaxation term in the continuity relation as in \eqref{U1currentbrokenconseq}, and allow for new terms in constitutive relations (see appendix for details), leading to a dispersion relation of the form
\begin{equation}
\omega = \pm \frac{f_s}{M_0} (q_o^2 + q^2) - i (\Gamma + D_o q^2 + \gamma q^4) + \cdots\, .
\end{equation}
Explicit breaking of the symmetry introduced three new parameters: $\Gamma,\,D_o,\,q_o$.
However, imposing that the hydrodynamic equations of motion be local shows that only two are independent:
\begin{equation}
\Gamma = q_o^2 \left(D_o - \gamma q_o^2\right)\,,
\end{equation}
which is the analogue to \eqref{eq_relation} in this case.
%

\strut

\begin{acknowledgments}

\paragraph{Note added:} After this work was posted on the arXiv, \cite{Armas:2021vku} appeared with a derivation of \eqref{eq_relation} using positivity of entropy production. They also report that extra coefficients arise when coupling to external sources. These do not affect our main result \eqref{eq_relation}. We give more details in the appendix.

We thank Akash Jain, Karl Landsteiner, Joaquin Rodriguez Nieva and Derek Teaney for helpful discussions. We thank Paolo Glorioso and Sean Hartnoll for insightful comments on a first version of this manuscript. We thank the authors of \cite{Armas:2021vku} for sharing with us the results of their work and pointing out some typos on a previous version of this work.
LVD is supported by the Swiss National Science Foundation and the Robert R. McCormick Postdoctoral Fellowship of the Enrico Fermi Institute. LVD also gratefully acknowledges support from the Simons Center for Geometry and Physics, Stony Brook University at which some of the research for this paper was performed. BG and VZ are supported by the European Research Council (ERC) under the European Union's Horizon 2020 research and innovation program (grant agreement No758759). BG would like to thank the Instituto de F\'isica Te\'orica, Universidad Aut\'onoma de Madrid and CSIC, where some of this work was carried out.
\end{acknowledgments}

\bibliographystyle{ourbst}
\bibliography{bibliography}

\onecolumngrid

\appendix 

\section{Constitutive relations with sources}\label{app_sources}

In this appendix, we review how hydrodynamic equations of motion as in Eq.~\eqref{eq_eom0} take the form \eqref{eq_eom} when the system is coupled to sources. This material is standard \cite{kadanoff1963hydrodynamic,chaikin1995principles}; we include it here for completeness because the appearance of sources in hydrodynamic equations of motion plays a key role in our derivation of the relation \eqref{eq_relation}.

Knowledge of the hydrodynamic equation of motion in the absence of sources
\begin{equation}\label{eq_eom0_app}
\dot n_a(q,t) + M_{ab}(q)n_b(q,t) = 0\, , 
\end{equation}
together with equilibrium thermodynamics data, in the form of the response of the densities to {\em static} sources $\delta\mu_a(x,t) = \delta \mu_a(x)$,
\begin{align}\label{eq_static_response}
\qquad\qquad &&
\delta n_a(q)
	= \chi_{ab}(q) \delta \mu_b(q)\, ,
	&&\hbox{(static response)}
\end{align}
is sufficient to determine the response to an arbitrary source $\delta \mu_a(x,t)$. Indeed, first notice that one can find the density response to an adiabatically activated source, switched off at $t=0$:
\begin{equation}\label{eq_adiabatic}
\delta \mu_a(t,q) = \delta \mu_a (q) \theta(-t) e^{0^+ t} .
\end{equation}
For $t\leq 0$, the density is given by the static response \eqref{eq_static_response}; for $t>0$, the sources vanish so the density can be obtained from the equation of motion without sources \eqref{eq_eom0_app}. Now since functions \eqref{eq_adiabatic} switched off at various times form a basis of functions, and we are working within linear response, we can determine the response to sources with any functional form $\delta \mu_a(x,t)$. 

Let us spell out how to do this in practice. For sources \eqref{eq_adiabatic}, the Fourier transform of the density $n_a(\omega,q) \equiv \int dt \, e^{-i\omega t}n_a(t,q)$ is given by
\begin{equation}\label{eq_dn_wq}
\begin{split}
\delta n_a(\omega,q) 
	&= \int_{-\infty}^0 dt \, e^{-i(\omega+i0^+) t} \chi_{ab}(q)\delta \mu_b(q) + \int_{0}^\infty dt \, e^{-i\omega t} \delta n_a (t,q)  \\
	&= \frac{\chi_{ab}(q)\delta \mu_b(q)}{-i\omega + 0^{+}} + \left(i\omega + M(q)\right)^{-1}_{ab} \chi_{bc}(q)\delta \mu_c(q) \, .
\end{split}
\end{equation}
To evaluate the second integral, we used (dropping indices)
\begin{equation}
\begin{split}
(i\omega + M) \int_0^{\infty}dt\, e^{-i\omega t} \delta n(t)
	&= \int_0^{\infty} \left((-\partial_t + M)e^{-i\omega t}\right) \delta n(t)\\
	&= \delta n(t=0) + \int_0^{\infty} e^{-i\omega t} (\partial_t + M)\delta n(t)\, , 
\end{split}
\end{equation}
and the last factor vanishes by the equations of motion \eqref{eq_eom0_app}. Writing \eqref{eq_dn_wq} in terms of the Fourier transform of the source \eqref{eq_adiabatic}
\begin{equation}
\delta \mu_a(\omega,q)
	= \int dt \, e^{-i\omega t } \delta \mu_a (t,q)
	= \frac{\delta \mu_a(q)}{-i\omega + 0^{+}} \, ,
\end{equation}
gives
\begin{equation}
i\omega \delta n_{a}(\omega,q) + M_{ab}(q) \left(\delta n_b(\omega,q) - \chi_{bc}(q)\delta \mu_c(\omega,q)\right)
	= 0 \, .
\end{equation}
Inverse Fourier transform in time finally yields
\begin{equation}\label{eq_eom_app}
\dot \delta n_{a}(t,q) + M_{ab}(q) \left(\delta n_b(t,q) - \chi_{bc}(q)\delta \mu_c(t,q)\right)
	= 0\, .
\end{equation}
While we derived this equation for sources taking the form \eqref{eq_adiabatic}, any source $\delta \mu_{a}(x,t)$ can be obtained from a linear combination of these and Eq.~\eqref{eq_eom_app} is linear in the source; Eq.~\eqref{eq_eom_app}  therefore describes the linear density response to arbitrary sources. This is equation \eqref{eq_eom} which we wished to prove.

This establishes how external sources enter hydrodynamic equations of motion. In practice, these may enter at the level of constitutive relations, or conservation laws (as in the case of anomalies \cite{Son:2009tf}, or for non-abelian symmetries, see e.g.~\eqref{eq_cons_FM}).

\section{Schwinger-Keldysh Effective Field Theory}\label{app_SK}

In the Schwinger-Keldysh approach to computing real time observables in a thermal state, operators automatically satisfy local constitutive relations in terms of the sources, so that relations of the form \eqref{eq_relation} are obtained effortlessly. We show this in this appendix, referring the reader to \cite{kamenev2011field} for more details on the method.%
	\footnote{Fluctuating hydrodynamics has recently been recast as an effective field theory on a Schwinger-Keldysh contour \cite{Grozdanov:2013dba,Haehl:2015pja,Crossley:2015evo,Jensen:2018hse,Chen-Lin:2018kfl} (see \cite{Glorioso:2018wxw} for a review).}

We consider the simplest example of a pseudo-Goldstone boson by studying a system in a phase with a spontaneously broken $U(1)$ symmetry, ignoring other conserved quantities such as heat. At low energies, the system is described by an effective field theory of a Goldstone boson $\phi$. Real time dynamics in a state described by a density matrix (which we will take to be thermal $\rho = e^{-H/T}/\Tr e^{-H/T}$) can be studied by placing the effective field theory on a closed-time path, or Schwinger-Keldysh contour; we denote the field on each leg of the contour by $\phi_1$ and $\phi_2$. We follow the notation in \cite{Glorioso:2018wxw} and define
\begin{equation}\label{eq:ra_basis}
\phi_r = \frac12 \left(\phi_1 + \phi_2 \right)\, , \qquad
\phi_a = \phi_1 - \phi_2\, . 
\end{equation}
We ignore interactions and only consider the most general quadratic action for $\phi$, in a gradient expansion. The shift symmetry $\phi_{1,2}\to \phi_{1,2} + a_{1,2}$ implies that $\phi$ must enter the action with a derivative. To leading and subleading order in derivatives, the action must take the form
\begin{equation}\label{eq_S0_SK}
\begin{split}
S_0 = \chi_{nn}&\int 
	\left(\dot \phi_a \dot \phi_r - c_s^2 \nabla\phi_a \nabla \phi_r\right)\\
	&+ \left(D_n\nabla^2 \phi_a \dot \phi_r + \frac{D_\phi}{c_s^2} \ddot \phi_a \dot \phi_r\right)
	+ iT \left(D_n(\nabla \phi_a)^2 + \frac{D_\phi}{c_s^2}\dot \phi_a^2\right) + \cdots\, 
\end{split}
\end{equation}
where $\chi_{nn},\,c_s,\,D_n,\,D_\phi$ are Wilsonian coefficients. Scaling $\omega\sim k$ and $\phi_a\sim \partial\phi_r = O(\partial^0)$, the terms in the first line are $O(\partial^0)$ and are non-dissipative, whereas the terms in the second line are $O(\partial)$ and dissipative. The KMS relation (or fluctuation-dissipation relation) fixes the coefficients of the two $\phi_a^2$ terms in terms of $D_n,\,D_\phi$. Finally, $\phi_r^2$ terms are forbidden in the action to guarantee that $\langle \phi_a(t) \phi_a(t')\rangle = 0$, see \cite{Glorioso:2018wxw} for more details. 

The retarded Greens function can be directly obtained from this Gaussian action:\footnote{The retarded Green's function is defined as $G^R_{AB}(t)\equiv i\theta(t) \langle[A(t),B]\rangle$, and $G^R(\omega) = \int dt\,e^{i\omega t} G^R(t)$. With this sign convention, $\omega \Im G^R(\omega) \geq 0$.}
\begin{equation}\label{eq_Gff_app_SK}
G^R_{\phi\phi}(\omega,q)
	= \langle \phi_r\phi_a\rangle(\omega,q)
	\simeq \frac{1}{f_s} \frac{c_s^2}{c_s^2 q^2 - \omega^2 - i D_n \omega q^2 - i \frac{D_\phi}{c_s^2} \omega^3}\, , 
\end{equation}
with $f_s \equiv \chi_{nn}c_s^2$. Expanding in poles and residues, one has
\begin{equation}\label{eq_Gphiphi_SK}
\begin{split}
G^R_{\phi\phi}(\omega,q)
	&= 
	\frac{\frac{c_s}{2|q|} + \frac{i}2 D_\phi + O(q)}{\omega + c_s |q| + \frac{i}2(D_n+D_\phi)q^2 + O(q^3)}
	-
	\frac{\frac{c_s}{2|q|} - \frac{i}2 D_\phi  + O(q)}{\omega - c_s |q| + \frac{i}2(D_n+D_\phi)q^2 + O(q^3)}\!\\
	&- \frac{iD_\phi}{\omega - i \frac{c_s^2}{D_\phi} + O(q^2)}\, .
\end{split}
\end{equation}
The last term is never important in the hydrodynamic regime and can be ignored (moreover, the apparent pole at the EFT cutoff is of course not a reliable prediction of the EFT).
One can compare this to the Green's function obtained from the Kadanoff-Martin approach, using $G^R = (-i\omega + M)^{-1}M\chi$, where $\chi$ and $M$ are given by \eqref{eq_chi_superfluid} and \eqref{eq_Mmatrix_superfluid}:
\begin{equation}\label{eq_Gphiphi}
G_{\phi\phi}^R(\omega,q)
	\simeq \frac{1}{f_s } \frac{c_s^2 - i D_\phi \omega + D_n D_\phi q^2}{c_s^2 q^2 +(D_nq^2 - i \omega)(D_\phi q^2 - i\omega)}\, .
\end{equation}
The expression \eqref{eq_Gphiphi_SK} agrees with \eqref{eq_Gphiphi}, up to the $O(q)$ and $O(q^3)$ terms in the numerator and denominator respectively, which can be accounted for with higher gradient terms in the action \eqref{eq_S0_SK}.

Breaking the symmetry allows for new terms in the action:
\begin{equation}
\label{eq_S1_SK}
S_1 = -\chi_{nn} \int \tilde\omega_o^2 \phi_a \phi_r + \tilde \Gamma \phi_a \dot \phi_r - iT \tilde \Gamma \phi_a^2\, , 
	\qquad\qquad
S= S_0 + S_1\, .
\end{equation}
The first is non-dissipative, and the last two -- whose coefficients are again related by KMS -- are dissipative. Note that breaking the symmetry has only introduced two new parameters: $\tilde \omega_o$ and $\tilde \Gamma$. In contrast, the traditional hydrodynamic approach produces three parameters $\omega_o,\,\Gamma$ and $\Omega$, which are then found to be related by imposing locality as discussed in the main text. The relation \eqref{eq_relation} is automatic in this effective field theory approach. Indeed, the $q=0$ retarded Green's function is now
\begin{equation}
\label{eq_g_relaxed}
G^R_{\phi\phi}(\omega,q=0)
	\simeq \frac{1}{f_s} \frac{c_s^2}{\tilde \omega_o^2 - \omega^2 - i\omega\tilde \Gamma - i \frac{D_\phi}{c_s^2}\omega^3}\, .
\end{equation}
Expanding in poles and residues as before, one finds that it agrees with what is found from the Kadanoff-Martin approach, namely
\begin{equation}\label{eq_Gphiphi_relax}
G_{\phi\phi}^R(\omega,q)
	\simeq \frac{1}{f_s q_o^2} \frac{\omega_o^2 + (\Gamma - i\omega)\Omega}{\omega_o^2 +(\Gamma - i \omega)(\Omega - i\omega)}\, .
\end{equation}
with the identification%
	\footnote{In finding this relation, we have assumed for simplicity that $\tilde \Gamma= O(\omega_o^2)$. One can lift this assumption, but the relation between the EFT parameters $\tilde \omega_o,\,\tilde \Gamma$ and the parameters appearing in the constitutive relation $\omega_o,\,\Gamma$ becomes more complicated.} 
\begin{equation}
\label{OmegarelSK}
\omega_o = \tilde \omega_o\, , \qquad
\Gamma = \tilde \Gamma\, , \qquad
\Omega = \omega_o^2 \frac{D_\phi}{c_s^2}\, .
\end{equation}
We therefore find that the relation \eqref{eq_relation} directly follows from the effective field theory approach to hydrodynamics.\footnote{In the works \cite{Baggioli:2020nay,Baggioli:2020haa}, while the symmetry-breaking terms in \eqref{eq_S1_SK} are also considered, the all-important $D_\phi$ term in \eqref{eq_S0_SK} is not included and hence their derivation of \eqref{OmegarelSK} is not complete.} This is thanks to the fact that this formulation is manifestly local, up to scales set by the UV cutoff for hydrodynamics, rather than up to the (parametrically smaller) thermal mass $\sim \omega_o$.

To understand better the microscopic origin of the symmetry-breaking terms in the effective action \eqref{eq_S1_SK}, consider deforming the original action \eqref{eq_S0_SK} by 
\begin{equation}\label{eq_def}
\delta S = \delta H
	= g  \int \mathcal{O}_Q\, ,
\end{equation}
where $\mathcal{O}_Q$ is an operator of charge $Q$ and $g \ll 1$. We then have
\begin{equation}
\dot N = i[N,\delta H] = g Q \int \mathcal{O}_Q\, ,
\end{equation}
To leading order in fields, \eqref{eq_def} can be written
\begin{equation}\label{eq_def_2}
\delta S = g \int \mathcal{O}^i_Q \left(e^{-iQ\phi_i}  + i Q \phi_i + O(\phi_i^2)\right)
	\to ig Q \int \mathcal{O}_Q^i \phi_i\, ,
\end{equation}
where in the last step we dropped the term $\mathcal{O}^i_Q e^{-iQ\phi_i}$ because it is invariant under the symmetries. Integrating out the microscopic degrees of freedom, this leads to a change in the effective action
\begin{equation}\label{eq_def_eff}
\delta S_{\rm eff}
	= \frac{(g Q)^2}{2}\int_{\omega,k} \phi_i \phi_j \langle \mathcal{O}_Q^i \mathcal{O}_Q^j\rangle(\omega,k)
\end{equation}
Let us focus on the most relevant deformation and take $k,\omega\to 0$. Now note that 
\begin{equation}
\langle \mathcal{O}_a \mathcal{O}_r\rangle = G^R_{\mathcal{O}_Q \mathcal{O}_Q}\, , \qquad
\langle \mathcal{O}_r \mathcal{O}_a\rangle = G^A_{\mathcal{O}_Q \mathcal{O}_Q}\, , \qquad
\langle \mathcal{O}_r \mathcal{O}_r\rangle = G_{\mathcal{O}_Q\mathcal{O}_Q} \simeq \frac{2T}{\omega} \Im G^R_{\mathcal{O}_Q\mathcal{O}_Q}\, , \qquad
\langle \mathcal{O}_a \mathcal{O}_a\rangle = 0\, ,
\end{equation}
where we used the fluctuation-dissipation theorem for the microscopic degrees of freedom.

The full quadratic action with $q\to 0$ is
\begin{equation}
\begin{split}
\frac{1}{\chi_{nn}}\mathcal L
	&= \dot \phi_r \dot \phi_a + \frac{(g Q)^2}{2\chi_{nn}} \left(G^R_{\mathcal{O}_Q\mathcal{O}_Q}(\omega) +G^A_{\mathcal{O}_Q\mathcal{O}_Q}(\omega)\right) \phi_r \phi_a  + i\frac{(g Q)^2}{2\chi_{nn}}G_{\mathcal{O}_Q\mathcal{O}_Q}(\omega) \phi_a \phi_a\\
	&\simeq
	\dot \phi_r \dot \phi_a - \tilde\omega_o^2 \phi_r \phi_a   - \tilde \Gamma \dot\phi_r \phi_a + iT\tilde \Gamma \phi_a^2\, , 
\end{split}
\end{equation}
where in the second line we expanded in $\omega$, used $G^R(\omega,0)= G^A(-\omega,0)$ and the fluctuation-dissipation theorem, and defined
\begin{equation}
\tilde\Gamma = -\frac{(g Q)^2}{\chi_{nn}}\lim_{\omega\to 0} \frac{1}{\omega} \Im G^R_{\mathcal{O}_Q\mathcal{O}_Q}(\omega)\, ,\qquad
\tilde\omega_o^2
	= -\frac{(g Q)^2}{\chi_{nn}} \lim_{\omega\to 0} \Re G^R_{\mathcal{O}_Q \mathcal{O}_Q}(\omega)\, .
\end{equation}

\subsection{Superfluid Frame Transformations}\label{app_frame}

As already mentioned, the Schwinger-Keldysh formulation of hydrodynamics systematically produces local constitutive relations for the conserved currents, as well as local conservation equations, to all orders in the hydrodynamic expansion and including fluctuations. However, frame in which we land by construction is somehow unusual from the point of view of hydrodynamics. In the previous appendix we directly integrated out the Goldstone field in order to obtain its Green's function. Here we take a step back and discuss the relevant hydrodynamic currents and equations of motion; this way we gain a slightly more general perspective on our derivation of Goldstone damping.

Let us consider again the SK action \eqref{eq_S0_SK}  for the superfluid, ignoring the quadratic in $\phi_{a}$, stochastic terms. The Noether procedure for the global shift symmetry $\phi_a\to\phi_a+c$ leads to the existence of a $U(1)$ current, whose off-shell form is
\begin{equation}\label{noether_currents_off}
	\begin{split}
		\hat{J}_r^{0} &= -\chi_{nn}\left(\dot\phi -\frac{D_\phi}{c_s^2} \ddot \phi\right)\,,\\
		\hat{J}_r^{i} &= \chi_{nn}\left(c_s^2 \partial^i\phi +D_n  \partial^i \dot \phi\right)\,,
	\end{split}
\end{equation}
omitting the $r$ indices. The equation of motion for $\phi_a$ is the current conservation
\begin{equation}
\partial_0 \hat {J}^{0} + \partial_i \hat {J}^{i} = 0\,,
\end{equation}
giving the $\phi-\phi$ Green's function \eqref{eq_Gff_app_SK}, from which one can then compute the Green's functions of the currents.

The transformation of $\phi$ under the shift symmetry fixes its relation to the chemical potential \cite{Crossley:2015evo}, i.e. the Josephson relation, to be
\begin{equation}\label{mdiss0}
\mu = -\dot\phi \,,
\end{equation}
to all orders in the derivative expansion. We thus obtain
\begin{equation}\label{sr_currents_mdiss}
	\begin{split}
		\hat{J}_r^{0} &= \chi_{nn}\left(\mu -\frac{D_\phi}{c_s^2} \dot \mu\right)\,,\\
		\hat{J}_r^{i} &= \chi_{nn}\left(c_s^2 \xi^i -D_n  \partial^i \mu\right)\,,
	\end{split}
\end{equation}
where we also identified the superfluid velocity $\xi^{i}=-\partial^i\phi$. This is the $\mu_{diss}=0$ frame of \cite{Bhattacharya:2011eea}, and is also the frame in which one naturally lands in in holography \cite{Donos:2021pkk}.

By performing a frame transformation
\begin{equation}\label{mtransverse}
\mu=-\dot\phi +\frac{D_\phi}{c_s^2} \ddot \phi\,,
\end{equation}
we arrive at the conventional form of the constitutive relations
\begin{equation}\label{sr_currents_transverse}
\begin{split}
	\hat{J}_r^{0} &= \chi_{nn} \mu\,,\\
	\hat{J}_r^{i} &= \chi_{nn}\left(c_s^2 \xi^i -D_n  \partial^i \mu\right)\,,
\end{split}
\end{equation}
to second order in derivatives.

Let us finally consider the effects of pinning. Adding back the action \eqref{eq_S1_SK} does not affect most of the above analysis; in particular, the off-shell form of the currents remains the same.\footnote{This can be seen by coupling the dynamical fields $\phi_{r,a}$ to sources in gauge invariant combinations as in the normal fluid case \cite{Crossley:2015evo}.} 
However, the currents are not conserved, and the equations of motion acquire the extra terms
\begin{equation}\label{eq_eoms_pinned}
	\partial_0 \hat {J}^{0} + \partial_i \hat {J}^{i} = \chi_{nn}\tilde\omega_o^2 \phi + \tilde \Gamma\chi_{nn} \dot\phi\,,
\end{equation}
which to ideal order give
\begin{equation}\label{sr_eom_ideal}
\dot\mu + c_s^2\partial_i \xi^{i} =\omega_o^2 \phi + \Gamma \dot\phi+\cdots\,.
\end{equation}
We can use \eqref{sr_eom_ideal} in order to put the Josephson relation and the constitutive relations partially on-shell.
In the conventional frame \eqref{mtransverse} this gives rise to the Goldstone damping  term in the Josephson relation
\begin{equation}\label{transverse_pr}
	\mu=-\dot\phi -\frac{D_\phi}{c_s^2} \omega_o^2 \phi +\cdots\,,
\end{equation}
and we recover the expressions used in the main text. However, one can also use the alternative frame \eqref{mdiss0}, in which the time component of the current becomes
\begin{equation}\label{sr_currents_pr}
J_r^{0} = \chi_{nn}\left(\mu -\frac{D_\phi}{c_s^2} \omega_o^2 \phi\right)+\cdots\,.
\end{equation}
We thus see that, when we go on-shell, the Goldstone damping term can be moved around among the constitutive relations and the Josephson equation by making frame transformations. Of course, the physical information resides in the retarded Green's functions, which remain invariant.

\section{Coupling to external sources}

After this paper appeared on the arXiv, \cite{Armas:2021vku} also appeared, deriving \eqref{eq_relation} and finding extra coefficients in the presence of external sources. In this appendix we clarify their origin in our formalism and show that they do not affect our results.

Let us consider coupling the SK effective field theory to external gauge fields $A_{r\mu},A_{a\mu}$ in the retarded-advanced basis \eqref{eq:ra_basis}. The superfluid Lagrangian \eqref{eq_S0_SK} becomes
\begin{equation}\label{eq_L0_SK}
	\mathcal{L}_0 = -\chi_{nn} \left[ -B_{a0} B_{r0} +\frac{D_\phi}{c_s^2} B_{a0} \dot B_{r0} + c_s^2 B_{ai} B_{ri} +D_n B_{ai} \dot B_{ri} +\cdots\right]\,,
\end{equation} 
where
\begin{equation}\label{eq_Bmu_def}
	B_{r\mu} = A_{r\mu} +\partial_\mu \phi_r \,, \qquad B_{a\mu} = A_{a\mu} +\partial_\mu \phi_a\,,
\end{equation}
and in \eqref{eq_L0_SK} we have imposed isotropy, dynamical KMS symmetry \cite{Glorioso:2017fpd,Glorioso:2018wxw}, and the Onsager relations. The dots contain higher derivative and stochastic terms quadratic in the $a$-fields.

Breaking the symmetry allows for the following new terms in the Lagrangian 
\begin{equation}\label{eq_L1_SK}
		\mathcal{L}_1 = -\chi_{nn} \left[\omega_o^2 \phi_a \phi_r + \Gamma \phi_a \dot \phi_r - \sigma \left(\phi_{a} \dot B_{r0} - B_{a0} \dot\phi_{r}\right) - c_s^2\kappa \left(\phi_{a} \partial_i B_{ri} - B_{ai} \partial_i\phi_{r}\right) \right]\,,
\end{equation}
after imposing dynamical KMS symmetry and the Onsager relations. 

The off-shell currents $\hat{J}^{\mu} \equiv -\delta\mathcal{L}/\delta B_{a\mu}$ now take the form
\begin{equation}\label{eq_curr_KMSOns}
	\begin{split}
		\hat{J}^{0} &= \chi_{nn}\left(-B_{0} +\frac{D_\phi}{c_s^2} \dot B_0 +\sigma \dot\phi \right)\,,\\
		\hat{J}^{i} &= \chi_{nn}\left(c_s^2 B^i +D_n \left( \partial^i B_0 +E^i \right) +c_s^2\kappa \partial^i \phi\right)\,,
	\end{split}
\end{equation}
and the equation of motion is
\begin{equation}\label{eq_eom_KMSOns}
	\partial_0 \hat {J}^{0} + \partial_i \hat {J}^{i}  = \chi_{nn}\left(\omega_o^2 \phi + \Gamma \dot \phi -\sigma \dot B_{0} - c_s^2\kappa \partial_i B_{i}\right)\,.
\end{equation}

Let us now go to the natural frame in which $J^0=\chi_{nn}\mu$ and put the currents partially on-shell, similarly to the previous appendix. We obtain the Josephson relation
\begin{equation}\label{eq_Josephson}
	\mu=-B_{0} +D_\phi \partial_i B^i -\frac{D_\phi}{c_s^2}\omega_o^2\phi +\sigma \dot\phi\,,
\end{equation}
while the currents take the more conventional form 
\begin{equation}\label{sr_currents}
	\begin{split}
		\hat{J}^{0} &= \chi_{nn} \mu\,,\\
		\hat{J}^{i} &= \chi_{nn}\left(c_s^2 B^i -D_n \left( \partial^i \mu -E^i \right) +c_s^2\kappa \partial^i \phi\right)\,.
	\end{split}
\end{equation}
to second order in derivatives. One can check that \eqref{eq_Josephson}, \eqref{sr_currents} and \eqref{eq_eom_KMSOns} are equivalent to the corresponding results of \cite{Armas:2021vku} in the presence of external fields, with $\sigma$ and $\kappa$ corresponding to $\sigma_\times$ and $\bar{f}_s$ in their notation.

Note that $\sigma$ and $\kappa$ do not affect the physical modes or the main result of our paper \eqref{eq_relation}. In particular, turning the background fields off, the dispersion relation can be written as
\begin{equation}
(1+\sigma)^2\omega^2 -\omega_o^2 -c_s^2(1+\kappa)^2q^2 +i(\Gamma +D_n q^2)\omega +i \frac{D_\phi}{c_s^2}\omega^3 =0\,,
\end{equation}
up to the order we are working at. We see that $\sigma$ and $\kappa$ enter as pinning corrections to the susceptibilities $\chi_{nn}$ and $c_s^2$. Of course, the susceptibilities will in general involve other pinning corrections too, appearing as corrections to the coefficients of the original action \eqref{eq_L0_SK}. In our paper we are consistently ignoring all such corrections, since we are only concerned with the leading order result \eqref{eq_relation}.

\section{Wigner crystal hydrodynamics \label{app:WC}}

In this appendix we summarize the hydrodynamics of (pinned) Wigner crystals. We refer to eg \cite{chaikin1995principles,Delacretaz:2017zxd} for more details, see also \cite{Armas:2019sbe,Armas:2020bmo}.

The free energy density for an isotropic Wigner crystal in mechanical equilibrium in two spatial dimensions is only a function of the symmetric, linear strain $u_{ij}=\nabla_{(i}u_{j)}$
\begin{equation}\label{f_wc}
f=\frac{B}2 \left(u_l^l\right)^2+G\left(u_{ij}u^{ij}-\frac12\left(u_l^l\right)^2\right)+\frac12G q_o^2 u_i u^i\,,
\end{equation}
where lower-case latin indices run over the spatial dimensions and indices are raised and lowered with the Kronecker delta $\delta_{ij}$. We have also added a small mass term which explicitly breaks translation symmetry, realized here as a constant shift of the Goldstone fields, but neglected for a moment couplings to charge and entropy fluctuations.

Turning off temporarily the mass term $q_o$, the field $h_{ij}$ conjugate to $u_{ij}$ is defined from the free energy as
\begin{equation}
    \left.h_{ij}\right|_{\delta\mu,\delta T=0}\equiv\frac{\delta f}{\delta u_{ij}}=B u_{l}^l\delta_{ij}+G\left(2u_{ij}-u_l^l\delta_{ij}\right)\,.
\end{equation}

In the presence of a mass, we should instead treat the displacements $u_i$ as the fundamental fields. Hence we define a new conjugate field
\begin{equation}
\label{hsourceWCpinned}
    h_i\equiv\frac{\delta f}{\delta u_{i}}=G q_o^2 u_i-\nabla^j h_{ij}\,.
\end{equation}

In Fourier space, the free energy takes the form
\begin{equation}
    f=\frac12\left((B+G)+G\frac{q_o^2}{q^2}\right)\lambda_\parallel(-\omega,-q) \lambda_\parallel(\omega,q)+\frac{G}2\left(1+\frac{q_o^2}{q^2}\right)\lambda_\perp(\omega,q)\lambda_\perp(-\omega,-q)\,,
\end{equation}
where we defined the longitudinal and transverse strains
\begin{equation}
    \lambda_\parallel=\nabla_i u^i\,,\qquad \lambda_\perp=\epsilon_{ij}\nabla^i u^j\,.
\end{equation}
The corresponding conjugate fields are denoted by $h_\parallel=i q_i h^{i}/q^2$, $h_\perp=-\epsilon^{ij}q_j  h_{i}/q^2$, or in real space $h_\parallel=-\nabla^i h_{i}/\nabla^2$, $h_\perp=\epsilon^{ij}\nabla_j h_{i}/\nabla^2$.

The conservation equations are 
\begin{align}\label{app_WC_conservationeqs}
	\dot n	+\nabla\cdot j =0\,, \quad	\dot s	+\nabla\cdot(j_Q/T) =0\,, \quad   \partial_t \pi^i+\nabla_j\tau^{ji}=-G q_o^2u^i-\Gamma \pi^i\,.
\end{align}
The momentum (non-)conservation equation picks up a nonzero right-hand side
with both a pinning term and a momentum relaxation rate. There are several ways to see how the pinning term arises. For instance, it is necessary for Onsager relations to hold. It also arises by computing $\partial_t\pi=i[H,\pi]$, approximating the effective Hamiltonian in the infrared by the free energy and remembering that $i[u^i(x),\pi^j(y)]=\delta(x-y)\left(\delta^{ij}+\partial^j u^i\right)$.

In the absence of pinning $q_o=0$, the susceptibility matrix has the form
\begin{equation}\label{susc_matrix_translation}
	\chi_0=  \left(\begin{array}{cccccc}
		\chi_{nn} &\chi_{ns} &0 &\chi_{n\lambda_\parallel} &0 &0 \\
		\chi_{ns} &\chi_{ss} &0 &\chi_{s\lambda_\parallel} &0 &0 \\
		0 &0 &\chi_{\pi\pi} &0 &0 &0 \\
		\chi_{n\lambda_\parallel} &\chi_{s\lambda_\parallel} &0 &\chi_{\lambda_\parallel\lambda_\parallel} &0 &0 \\
		0 &0 &0 &0 &\chi_{\pi\pi}&0 \\
		0 &0 &0 &0 &0 &\chi_{\lambda_\perp\lambda_\perp}
	\end{array}\right)\,,
\end{equation}
with
\begin{equation}
	\chi_{\lambda_\parallel\lambda_\parallel}=\frac{1}{B+G}\,,\qquad \chi_{\lambda_\perp\lambda_\perp}=\frac{1}{G}\,.
\end{equation}
We have also restored the off-diagonal couplings of the longitudinal Goldstone to charge and entropy, $\chi_{n\lambda_{\parallel}}$, $\chi_{s\lambda_{\parallel}}$, which are related to the thermal and mass expansion in elasticity literature. These couplings are essential to match to holographic models, \cite{Donos:2019hpp,Armas:2019sbe,Ammon:2020xyv}.

The most general constitutive relations for the currents take the form \cite{chaikin1995principles,Delacretaz:2017zxd,Armas:2019sbe,Armas:2020bmo}:\footnote{The coefficients of the $\nabla^i\mu,\nabla^i T$ terms in \eqref{WCconstrel}, \eqref{WCconstrel2} are chosen such that $\sigma_0,\alpha_0,\bar\alpha_0,\bar{\kappa}_0$ are given by simple Kubo formul\ae, eg $\sigma_o=\lim_{\omega\to0}\textrm{Im}G^R_{jj}(\omega,q=0)/\omega$.}
\begin{align}
	\label{WCconstrel}
	j^i &=nv^i -(\sigma_0-\chi_{n\lambda_\parallel}\gamma_{1l}(B+G))\nabla^i\mu -(\alpha_0-\chi_{s\lambda_\parallel}\gamma_{1l}(B+G))\nabla^i T -\gamma_{1l}(B+G)\nabla^i \lambda_{\parallel} -\gamma_{1t}G\epsilon^{ij}\nabla_j \lambda_{\perp}\,,\\
	\label{WCconstrel2}
	j_Q^i/T &=sv^i -(\bar\alpha_0-\chi_{n\lambda_\parallel}\gamma_{2l}(B+G))\nabla^i\mu -\left(\bar{\kappa}_0/T-\chi_{s\lambda_\parallel}\gamma_{2l}(B+G)\right)\nabla^i T -\gamma_{2l}(B+G)\nabla^i \lambda_{\parallel} -\gamma_{2t}G\epsilon^{ij}\nabla_j \lambda_{\perp}\,,\\
	\tau^{ij} &=\left(p- (B+G)\lambda_{\parallel}\right)\delta^{ij} - G\lambda_{\perp}\epsilon^{ij}  -\sigma^{ij}\,,
\end{align}
with 
\begin{align}
	p &=-\epsilon+T s+\mu n+v_i\pi^i\,, \\
	\sigma^{ij} &= \zeta \delta^{ij}\left(\nabla \cdot v\right) +\eta\left[2\nabla^{(i}v^{j)}-\delta^{ij}\left(\nabla \cdot v\right)\right]\,.
\end{align}
Positivity of entropy production in the presence of background sources can be used to show that $\gamma_{1t}=\gamma_{1l}$ and $\gamma_{2t}=\gamma_{2l}$, \cite{Armas:2019sbe,Armas:2021vku}. Since these terms are transverse, they do not enter the equations of motion and do not affect our analysis.\footnote{They were not written down in \cite{Delacretaz:2017zxd}.}

The Josephson relation for the Goldstones is
\begin{align}
 \partial_t u_i =v_i + \gamma_{3c} \nabla_i \mu +\gamma_{3h} \nabla_i T +\xi_\parallel(B+G) \nabla_{i} \lambda_{\parallel} +\xi_\perp G\varepsilon_{ij}\nabla^{j} \lambda_{\perp}\,.
\end{align}
Taking the divergence and curl of the Josephson relation then leads to
\begin{align}
	&\dot\lambda_{\parallel} -\nabla \cdot v - \gamma_{3c} \nabla^2 \mu - \gamma_{3h} \nabla^2 T -\xi_\parallel (B+G)\nabla^2 \lambda_{\parallel} =0\,, \\
	&\dot\lambda_{\perp} - \nabla \times v -\xi_\perp G\nabla^2 \lambda_{\perp} =0\,.
\end{align}

In terms of the fields $\{n,s,\pi_{\parallel},\lambda_{\parallel},\pi_{\perp},\lambda_{\perp}\}$, the conservation equations take the form \eqref{eq_eom0} with
\begin{equation}
	\label{MchiWC}
	M(q)\cdot\chi_0=  \left(\begin{array}{cccccc}
		\sigma_0 q^2 &\alpha_0 q^2 &iq n &\gamma_{1l}q^2 &0 &0 \\
		\bar\alpha_0 q^2 &\frac{\bar\kappa_0}{T} q^2 &iq s &\gamma_{2l}q^2 &0 &0 \\
		iq n &iq s &(\zeta+\eta)q^2 &-iq &0 &0 \\
		(\gamma_{3c}+\chi_{n\lambda_\parallel}\xi_\parallel (B+G)) q^2 &(\gamma_{3h} +\chi_{s\lambda_\parallel}\xi_\parallel (B+G))q^2 &-iq &\xi_\parallel q^2 &0 &0 \\
		0 &0 &0 &0 &\eta q^2 &-iq  \\
		0 &0 &0 &0 &-iq &\xi_\perp q^2 
	\end{array}\right)\,.
\end{equation}

The Onsager relations imply some constraints on the matrix $M\cdot\chi_{0}$ 
\begin{equation}\label{Onsager_Mchi}
	S \cdot\left(M(-q)\cdot\chi_0\right)^T = M(q)\cdot\chi_0\cdot S\,,
\end{equation}
with $S$ being the matrix of time-reversal eigenvalues of the corresponding fields. Here $S=\textrm{diag}(1,1,-1,1,-1,1)$ and then \eqref{Onsager_Mchi} gives 
\begin{equation}\label{Ons_1}
	\bar\alpha_0=\alpha_0\,,\quad \gamma_{1l}=\gamma_{3c}+\chi_{n\lambda_\parallel}\xi_\parallel (B+G)\,,\quad \gamma_{2l}=\gamma_{3h}+\chi_{s\lambda_\parallel}\xi_\parallel (B+G)\,.
\end{equation}
Isotropy of the retarded Green's functions in the $q\to0$ limit implies that 
\begin{equation}
\label{reldifflongperp}
	\xi_\parallel=\xi_\perp\equiv\xi\,.
\end{equation}
However, this can also be seen by demanding locality of the matrix $M\cdot\chi_0$ in the general basis $(u_i,u_j)$ (instead of diagonalizing along the longitudinal and transverse directions), which is only obtained when \eqref{reldifflongperp} holds (see the following appendix). 

Adding back pinning, the susceptibility matrix becomes\footnote{A new $\chi_{n u_x}$ susceptibility is in principle allowed, but vanishes for isotropic phases.}
\begin{equation}\label{deltachi_def}
    \chi=\left(\chi_0^{-1}+\Delta\chi^{-1}\right)^{-1}
\end{equation}
where
\begin{equation}\label{deltachi_translation}
	\Delta\chi^{-1}=  \left(\begin{array}{cccccc}
		0 &0 &0 &0 &0 &0 \\
		0 &0 &0 &0 &0 &0 \\
		0 &0 &0 &0 &0 &0 \\
		0 &0 &0 &G \frac{q_o^2}{q^2} &0 &0 \\
		0 &0 &0 &0 &0 &0 \\
		0 &0 &0 &0 &0 &G \frac{q_o^2}{q^2}
	\end{array}\right)\,,
\end{equation}
which is divergent as $q\to0$ because we are using the $\lambda_{\parallel},\lambda_{\perp}$ variables. Evaluating the matrix $\chi$, we find that non-local contributions proliferate in the entire coupled sector $(n,s,\lambda_\parallel)$. It is interesting to note that
\begin{equation}
\left(\chi\right)_{nn}=\chi_{nn}-\frac{G(B+G)q_o^2\chi_{n\lambda_{\parallel}}^2}{G q_o^2+(B+G)q^2}\quad\underset{q\to0}{\longrightarrow}\quad\chi_{nn}-(B+G)\chi_{n\lambda_\parallel}^2\,.    
\end{equation}
Here recall that $\chi_{nn}=\left(\chi_0\right)_{nn}$, the density-density static susceptibility in the absence of pinning. In the presence of pinning, there is an extra contribution to this static susceptibility coming from the nonzero off-diagonal matrix element with $\lambda_\parallel$. Similar expressions can be obtained for $\left(\chi\right)_{ns},\left(\chi\right)_{ss},\left(\chi\right)_{n\lambda_{\parallel}}$ and $\left(\chi\right)_{s\lambda_{\parallel}}$.

The constitutive relations (we remind the reader we only include linear terms) allow for several new dissipative terms in the presence of pinning
\begin{align}
	\label{WCconstrelpinned}
	j^i &=nv^i +\Omega_n u^i-\sigma_0\nabla^i\mu -\alpha_0\nabla^i T -\gamma_{1l}\nabla^i \tilde{h}_{\parallel}-\gamma_{1l}\epsilon^{ij}\nabla_j \tilde{h}_{\perp} \,,\\
	j_Q^i/T &=sv^i+\Omega_s u^i -\bar\alpha_0\nabla^i\mu -\bar{\kappa}_0/T\nabla^i T -\gamma_{2l}\nabla^i \tilde{h}_{\parallel}-\gamma_{2l}\epsilon^{ij}\nabla_j \tilde{h}_{\perp} \,,\\
	\tau^{ij} &=\left(p- (B+G)\lambda_{\parallel}\right)\delta^{ij} - G\lambda_{\perp}\epsilon^{ij}  -\sigma^{ij}\,,\\
 \dot u_i& =-\Omega u_i+v_i + \gamma_{3c} \nabla_i \mu +\gamma_{3h} \nabla_i T +\xi(B+G) \nabla_{i} \lambda_{\parallel} +\xi G\epsilon_{ij}\nabla^{j} \lambda_{\perp}\,,
\end{align}
where we defined
\begin{equation}\label{tildeh_def}
\tilde{h}_{\parallel} = (B+G) \left(\lambda_{\parallel} -\chi_{n\lambda_\parallel}\mu -\chi_{s\lambda_\parallel}T \right)\,, \quad	\tilde{h}_{\perp} = G \lambda_{\perp}\,.
\end{equation}
The new matrix $M\cdot\chi$ is local only if we impose
\begin{align}\label{Omega_wigner}
	\Omega= G q_o^2 \xi\,,\quad \Omega_n=G q_o^2\gamma_{1l}\,,\quad \Omega_s=G q_o^2\gamma_{2l}\,.
\end{align}
We observe that, as in the $U(1)$ case, it is possible to rewrite the dissipative terms  in the constitutive relations in terms of the conjugate fields:
\begin{align}
	\label{WCconstrelpinned_sources}
	j^i &=nv^i -\sigma_0\nabla^i\mu -\alpha_0\nabla^i T -\gamma_{1l}\nabla^i h_{\parallel}-\gamma_{1l}\epsilon^{ij}\nabla_j h_{\perp} \,,\\
	j_Q^i/T &=sv^i-\alpha_0\nabla^i\mu -\left(\bar{\kappa}_0/T\right)\nabla^i T -\gamma_{2l}\nabla^i h_{\parallel}-\gamma_{2l}\epsilon^{ij}\nabla_j h_{\perp} \,,\\
	\tau^{ij} &=\left(p- (B+G)\lambda_{\parallel}\right)\delta^{ij} - G\lambda_{\perp}\epsilon^{ij}  -\sigma^{ij}\,,\\
 \dot \lambda_\parallel& =\xi\nabla^2 h_\parallel+\nabla^iv_i + \gamma_{1l} \nabla^2 \mu +\gamma_{2l} \nabla^2 T \,,\\
   \dot \lambda_\perp& =\xi\nabla^2 h_\perp+\epsilon^{ij}\nabla_i v_j\,,
\end{align}
where we imposed the Onsager relations \eqref{Ons_1}.
	
The expression for the resistivity \eqref{resistivityWCrelaxed} in the main text can be derived using the standard Kadanoff-Martin procedure from the zero frequency limit of the density-density retarded Green's functions
\begin{equation}\label{DC_cond_WC}
\frac1{\rho_{dc}}=\sigma_{dc}= \lim_{\omega\to0,\,k\to0} \frac{i \omega}{k^2} G^{R}_{nn}(\omega,k) \simeq \sigma_0 + \frac{1}{\chi_{\pi\pi}}\frac{n^2 \Omega-2n\chi_{\pi\pi}\omega_o^2\gamma_{1l}-\Gamma\omega_o^2\chi_{\pi\pi}^2\gamma_{1l}^2}{\Omega\Gamma+\omega_o^2}\,,
\end{equation}
where $\omega_o^2\equiv G q_o^2/\chi_{\pi\pi}$. In a Galilean-invariant state, the charge density $n\mapsto ne$ with $n$ now the density, the momentum susceptibility is given by $\chi_{\pi\pi}=m^\star n$, with $m^\star$ the effective electronic mass, and the transport coefficients $\sigma_0$, $\alpha_0$ and $\gamma_{1l}$ vanish due to the Ward identity for Galilean boosts \cite{chaikin1995principles}, which enforces $j=e\pi/m^\star= ne v^i$. We have also reinstated the electron unit charge $e$ in \eqref{resistivityWCrelaxed}.

\subsection{Proving \eqref{reldifflongperp} \label{app:localitywcnopinning}}

Here we wish to show how \eqref{reldifflongperp} arises from locality. For simplicity, we turn off pinning and restrict the matrices to their $(u_x,u_y)$ elements, which are the only important ones for our argument. 
The susceptibility matrix is
\begin{equation}
    \chi=\left(\begin{array}{cc}
         \frac{G q_x^2+(B+G)q_y^2}{q^4 G(B+G)}&-\frac{q_x q_y B}{q^4 G(B+G)}  \\
         -\frac{q_x q_y B}{q^4 G(B+G)}& \frac{G q_y^2+(B+G)q_x^2}{q^4 G(B+G)}
    \end{array}\right)
\end{equation}
which was calculated by defining
\begin{equation}
h_i\equiv \frac{\delta f}{\delta u_i} =-\nabla^j h_{ij}   
\end{equation}
and then
\begin{equation}
    \chi^{-1}_{ij}=\frac{\delta h_i}{\delta u_j}
\end{equation}

Writing the Josephson relation as
\begin{align}
 \partial_t u_i =v_i + \gamma_{3c} \nabla_i \mu +\gamma_{3h} \nabla_i T +D_1 \nabla_i h_j^j+D_2 \nabla_j h_{i}^j  \,,
\end{align}
the matrix $M$ (restricted to $(u_x,u_y)$) is
\begin{equation}
    M=\left(\begin{array}{cc}
       2 B D_1 q_x^2+D_2(B q_x^2+G q^2)  & 2B D_1 q_x q_y+B D_2 q_x q_y \\
        2B D_1 q_x q_y+B D_2 q_x q_y & 2B D_1 q_y^2+D_2(B q_y^2+G q^2)
    \end{array}\right)
\end{equation}
which is local. Now we look at 
\begin{equation}
    M\cdot \chi=\left(\begin{array}{cc}
    D_1+\frac{2 D_1 B q_x^2}{q^2(B+G)}&\frac{2 D_1 B q_x q_y}{q^2(B+G)} \\
    \frac{2 D_1 B q_x q_y}{q^2(B+G)}&D_2 +\frac{2 D_1 B q_y^2}{q^2(B+G)}
    \end{array}\right)
\end{equation}
which is local only if $D_1=0$. Projecting along the longitudinal and transverse directions leads to our result \eqref{reldifflongperp}.

\section{Wigner crystal hydrodynamics in a magnetic field\label{app:WC_m}}

Here we outline the hydrodynamics of Wigner crystals in the presence of a magnetic field \cite{Delacretaz:2019wzh}, see also \cite{Baggioli:2020edn,Amoretti:2021fch}.

The free energy of the pinned magnetophonons is still given by \eqref{f_wc}, and the conservation equations take the form:
\begin{align}
	\dot n	+\nabla\cdot j =0\,, \quad	\dot s	+\nabla\cdot(j_Q/T) =0\,, \quad   \dot \pi^i+\partial_j\tau^{ji}= \hat{B} \epsilon^{ij} j_j+\Omega_\pi^{ij} u_j-\Gamma^{ij} \pi_j\,,
\end{align}
where $\hat{B}$ is the background magnetic field. The equation for momentum now includes a more general momentum relaxation matrix $\Gamma^{ij}\equiv\Gamma \delta^{ij}+\Gamma_H \hat{B} \epsilon^{ij}$, while the pinning term $\Omega_\pi^{ij}$ appears for the same reasons as in the Wigner crystal case without magnetic field (see the discussion below \eqref{app_WC_conservationeqs}). This quantity is fixed by the (non-)conservation equation (an expression will be given below in \eqref{Omega_wigner_m2}).

In the absence of pinning, the constitutive relations for the currents are given by 
\begin{align}\label{WCconstrel_m}
	j^i &=\gamma_{1c}^{ij}v_j -\sigma_0^{ij}\nabla_j\mu -\alpha_0^{ij}\nabla_j T -\gamma_{1l}^{ij}\nabla_j h_{\parallel} -\gamma_{1t}^{ij}\epsilon_{jk}\nabla^k h_{\perp}\,,\\
	j_Q^i/T &=\gamma_{2h}^{ij}v_j -\bar\alpha_0^{ij}\nabla_j\mu -(\bar{\kappa}_0^{ij}/T)\nabla_j T -\gamma_{2l}^{ij}\nabla_j h_{\parallel} -\gamma_{2t}^{ij}\epsilon_{jk}\nabla^k h_{\perp}\,,\\
	\tau^{ij} &=\left(p- (B+G)\lambda_{\parallel}\right)\delta^{ij} - G \lambda_{\perp}\epsilon^{ij}  -\sigma^{ij}\,.
\end{align}

The most general Josephson relation is
\begin{align}
	\dot u^i =\gamma_{3g}^{ij}v_j + \gamma_{3c}^{ij}\nabla_j\mu +\gamma_{3h}^{ij} \nabla_j T +\xi_\parallel^{ij}\nabla_j h_{\parallel} +\xi_\perp^{ij} \epsilon_{jk}\nabla^k h_{\perp}\,.
\end{align}
In the above expressions, all the matrix transport coefficients have the general form
\begin{equation}
\gamma^{ij}\equiv\gamma \delta^{ij}+\gamma_H \hat{B}\epsilon^{ij}\,.
\end{equation}
As before, the susceptibility matrix is given by \eqref{susc_matrix_translation}. The Onsager relations
\begin{equation}\label{Onsager_Mchi_m}
	S \left(M(-q,-\hat{B})\chi_0\right)^T = M(q,\hat{B})\chi_0 S\,,
\end{equation}
impose a set of constraints among the various transport coefficients\footnote{In contrast to the pure Wigner crystal case, now the Onsager relations are sufficient in order to show that $\gamma_{1l}^{ij}=\gamma_{1t}^{ij}$.}
\begin{align}\label{Ons_1_m}
	\bar\alpha_0=\alpha_0\,,\quad \gamma_{1c}&=n-\hat{B}^2\sigma_{0H}\,,\quad \gamma_{1l}=\gamma_{1t}=\gamma_{3c}\,,\quad  \gamma_{1lH}=\gamma_{1tH}=\gamma_{3cH}\,,\nonumber\\
	\gamma_{1cH}=\sigma_{0}\,,\quad \gamma_{2h}&=s-\hat{B}^2\alpha_{0H}\,,\quad \gamma_{2l}=\gamma_{3h}\,,\quad \gamma_{2hH}=\alpha_{0}\,,\quad \gamma_{2tH}=\gamma_{3hH}\,,\nonumber\\
	\xi_{\parallel H}=\xi_{t H}\,,\quad \gamma_{3g}&=1+\hat{B}^2\gamma_{3cH}\,,\quad \gamma_{3gH}=-\gamma_{3c}\,,\nonumber
\end{align} 
and then the conservation equations give
\begin{equation}
	\label{MchiWC_m}
	M\chi_0=  
	\resizebox{.9\hsize}{!}{$\left(\begin{array}{cccccc}
			\sigma_0 q^2 &\alpha_0 q^2 &iq\left(n-\hat{B}^2\sigma_{0H}\right) &\gamma_{3c}q^2 &iq \hat{B}\sigma_{0} &\hat{B}\gamma_{3cH}q^2 \\
			\alpha_0 q^2 &\frac{\bar\kappa_0}{T} q^2 &iq \left(s-\hat{B}^2\alpha_{0H}\right) &\gamma_{3h}q^2 &iq \hat{B}\alpha_{0} &\hat{B}\gamma_{3hH}q^2 \\
			iq\left(n-\hat{B}^2\sigma_{0H}\right) &iq\left(s-\hat{B}^2\alpha_{0H}\right) &(\zeta+\eta)q^2 +\hat{B}^2\sigma_{0} &-iq\left(1+\hat{B}^2\gamma_{3cH}\right) &-\hat{B}\left(n-\hat{B}^2\sigma_{0H}\right) &iq\hat{B}\gamma_{3c} \\
			\gamma_{3c} q^2 &\gamma_{3h} q^2 &-iq\left(1+\hat{B}^2\gamma_{3cH}\right) &\xi_\parallel q^2 &iq\hat{B}\gamma_{3c} &\hat{B}\xi_{\parallel H}q^2 \\
			-iq\hat{B}\sigma_{0} &-iq\hat{B}\alpha_{0} &\hat{B}\left(n-\hat{B}^2\sigma_{0H}\right) &-iq\hat{B}\gamma_{3c} &\eta q^2 +\hat{B}^2\sigma_{0} &-iq\left(1+\hat{B}^2\gamma_{3cH}\right)\\
			-\hat{B}\gamma_{3cH} q^2 &-\hat{B}\gamma_{3hH} q^2 &-iq\hat{B}\gamma_{3c} &-\hat{B}\xi_{\parallel H} q^2 &-iq\left(1+\hat{B}^2\gamma_{3cH}\right) &\xi_{\perp}q^2
		\end{array}\right)$}\,.
\end{equation}

The dispersion relations for the modes are generally rather complicated. However, we note that the magnetophonons generally disperse as $\omega\sim q^2-i q^2$, consistently with the observation in \cite{Baggioli:2020edn}.

Adding back pinning, the susceptibility matrix gets modified as in \eqref{deltachi_def},  \eqref{deltachi_translation}, and thus the constitutive relations take the form
\begin{align}\label{WCconstrelpinned_m}
j^i &=nv^i +\Omega_{n}^{ij}u_j-\sigma_0^{ij}\left(\hat{B}\epsilon_{jk}v^k+\nabla_j\mu\right) -\alpha_0^{ij}\nabla_j T -\gamma_{3c}^{ij}\left(\nabla_j \tilde{h}_{\parallel} +\epsilon_{jk}\nabla^k \tilde{h}_{\perp}\right)\,,\\
j_Q^i/T &=\left(s\delta^{i}_{j}-\alpha_0^{ik}\hat{B}\epsilon_{kj}\right) v^j +\Omega_{s}^{ij}u_j-\bar{\alpha}_0^{ij}\nabla_j\mu -(\bar{\kappa}_0^{ij}/T)\nabla_j T \nonumber\\
&-\left(\gamma_{3h}\delta^{ij}+\gamma_{2lH} \hat{B}\epsilon^{ij}\right)\nabla_j \tilde{h}_{\parallel} -\left(\gamma_{2t}\delta^{ij}+\gamma_{3cH} \hat{B}\epsilon^{ij}\right)\epsilon_{jk}\nabla^k \tilde{h}_{\perp}\,,\\
\tau^{ij} &=\left(p- \left(B+G\right)\lambda_{\parallel}\right)\delta^{ij} - G\lambda_{\perp}\epsilon^{ij}  -\sigma^{ij}\,,\\
\partial_t u^i &=v^i +\Omega^{ij}u_j+ \gamma_{3c}^{ij}\left(\hat{B}\epsilon_{jk}v^k+\nabla_j\mu\right) +\gamma_{3h}^{ij} \nabla_j T +\xi^{ij}\left(\nabla_j \tilde{h}_{\parallel} + \epsilon_{jk}\nabla^k \tilde{h}_{\perp}\right)\,,
\end{align}
where we used rotational invariance to set $\xi_\parallel=\xi_\perp\equiv\xi$, we defined
\begin{equation}
\xi^{ij} = \xi\delta^{ij}+\hat{B}\xi_{\parallel H}\epsilon^{ij}\,,
\end{equation}
and used the notation introduced in \eqref{tildeh_def}.

Locality of the new matrix $M\cdot\chi$ implies that
\begin{align}\label{Omega_wigner_m}
\Omega^{ij} = G q_o^2\xi^{ij}\,,
\end{align}
as well as
\begin{align}\label{Omega_wigner_m2}
\Omega_{n}^{ij} = G q_o^2\gamma_{3c}^{ij}\,,\quad \Omega_{s}^{ij} = G q_o^2\gamma_{3h}^{ij}\,,\quad \Omega_{\pi}^{ij} = G q_o^2 \left(\delta^{ij} + \hat{B}\gamma_{3c}^{ik}\epsilon_{k}{}^{j}\right)\,.
\end{align}
All in all, apart from the usual thermoelectric conductivities, the independent transport coefficients are $\gamma_{3c}\,,\gamma_{3cH}\,,\gamma_{3h}\,,\gamma_{3hH}\,,\xi\,,\xi_{\parallel H}\,,\gamma_{2t}\,,\gamma_{2lH}$. The first 6 of them determine the Goldstone damping rates and the contribution of the Goldstone modes to the currents.

The expression for the Goldstone relaxation rates in \eqref{Omega_wigner_m} was also obtained in \cite{Amoretti:2021fch,Amoretti:2021lll,Donos:2021ueh}, either through direct holographic calculation, or by consistency between the frequency-dependent correlators and the Ward identities in the hydrodynamic regime. The latter derivation does not make clear that relations of this type simply follow from locality. The Goldstone contributions in \eqref{Omega_wigner_m2} to the currents and the momentum equation were not explicitly identified in \cite{Amoretti:2021fch,Amoretti:2021lll}, whereas the comparison with \cite{Donos:2021ueh} is not straightforward, since there, by construction, the fluid velocity is integrated out and the Goldstone fields are partially on shell.

\section{Ferromagnets}

In this section, after reviewing the standard hydrodynamics of ferromagnets \cite{PhysRev.188.898}, we discuss weak explicit symmetry breaking and find new constraints on the resulting hydrodynamic theory. A ferromagnet is a system that spontaneously breaks $SU(2)$ spin symmetry down to $U(1)$, and has a finite magnetization
\begin{equation}\label{eq_condensate}
\langle n_a\rangle = \delta_a^3 M_0\, .
\end{equation}
In the presence of sources $A_\mu^a$ for the $SU(2)$ symmetry, the current is covariantly conserved
\begin{equation}\label{eq_cons_FM}
0
	= \nabla_\mu j^\mu_a
	= \partial_\mu j^\mu_a + f_{ab}{}^c A_\mu^b j^\mu_c\, .
\end{equation}
Because of the expectation value \eqref{eq_condensate}, the second term affects linearized hydrodynamics. Turning on only background chemical potentials $A^a_0 \equiv \mu^a$, the linearized continuity relation is
\begin{equation}\label{eq_cons_approx}
0 \simeq \partial_\mu j^\mu_a + \epsilon_{ab} \mu^b M_0\, .
\end{equation}
Since sources always appear in combination with densities in hydrodynamic equations (see Eq.~\eqref{eq_eom}), this will fix a term in the current constitutive relation. The free energy must take the form
\begin{equation}
F
	=\int \frac{1}{2} \frac{f_s}{M_0^2} (\nabla n_a)^2 + \frac{1}{2\chi} (n_3)^2\, ,  
\end{equation}
where $a=1,2$. These densities are Goldstones, which is why they must appear with gradients in the free energy.\footnote{Specifically, their linearized algebra $[n_a,n_b]\simeq i\epsilon_{ab}M_0$ implies that they are related to the Goldstones as $n_a = - M_0 \epsilon_{ab}\phi^b$ -- this justifies naming the coefficient above $f_s/M_0^2$, since in terms of Goldstones it appears as $F = \int \frac12 f_s (\nabla\phi)^2$.}

The current constitutive relations are
\begin{subequations}
\begin{align}
j_3^i
	&= - \tilde D\nabla^i n_3 + \cdots\, , \\
j_a^i
	&= \frac{f_s}{M_0} \epsilon_{ab}\nabla^i n^b +  \gamma\nabla^i \nabla^2 n_a + \cdots\, .
\end{align}
\end{subequations}
In the second line, the coefficient of the first term was fixed by requiring that the densities and external sources appear in the equation of motion \eqref{eq_cons_approx} as $n_a - \chi_{ab}\mu^b$. This same argument forbids a regular diffusive term $D\nabla^i n_a$, as it would require a nonlocal source term in the constitutive relation. Since $n_3$ decouples, let us focus on $n_a$, $a=1,2$. These satisfy the equation of motion 
\begin{equation}
0 = \dot n_a + \frac{f_s}{M_0} \epsilon_{ab}\nabla^2 n_b + \gamma (\nabla^2)^2 n_a + \cdots\, , 
\end{equation}
leading to modes
\begin{equation}
\omega = \pm \frac{f_s}{M_0} q^2 - i \gamma q^4\, .
\end{equation}

Let us now assume that the $SU(2)$ symmetry is weakly broken. In realistic experimental situations, the $SU(2)$ symmetry is explicitly broken by a combination of spin-orbit coupling, dipolar interactions, spatial anisotropies, etc., and the hydrodynamic description will depend on whether a $U(1)$ spin rotation remains approximately conserved, and if so what its orientation with respect to the order parameter is. Here for simplicity we will assume that the entire $SU(2)$ symmetry is broken without any preferred axis (e.g.~magnons can leak out of the system), and leave a more detailed analysis of other situations for future work. Explicit breaking of the symmetry allows for a small mass in the free energy
\begin{equation}
F
	=\int \frac{1}{2} \frac{f_s}{M_0^2} \left((\nabla n_a)^2 + q_o^2 (n_a)^2\right)\, .
\end{equation}
This also allows a relaxation term in the conservation law, along with a pinning term as in \eqref{app_WC_conservationeqs}
\begin{equation}
\nabla_\mu j^\mu_a = - \Gamma n_a + \frac{f_s}{M_0}q_o^2 \epsilon_{ab}n^b +\cdots\, .
\end{equation}
Finally, it may allow the diffusive term that was previously forbidden in the constitutive relation
\begin{equation}
j_a^i
	= \frac{f_s}{M_0} \epsilon_{ab}\nabla^i n^b - D_o \nabla^i n_a +  \gamma\nabla^i \nabla^2 n_a + \cdots\, .
\end{equation}
We have labelled the new coefficients as $q_o,\,\Gamma,\, D_o$. Demanding locality of the equations of motion in the presence of sources, one finds that these are not all independent, but must satisfy the relation
\begin{equation}
\Gamma = q_o^2 \left(D_o - \gamma q_o^2\right)
	\qquad \hbox{or} \qquad
	D_o = \gamma q_o^2 + \frac{\Gamma}{q_o^2}\, .
\end{equation}
The modes are now
\begin{equation}
\omega \simeq \pm \frac{f_s}{M_0} (q_o^2 + q^2)  - i \left(\Gamma + D_o q^2 + \gamma q^4\right) = (q_o^2 + q^2) \left[\pm \frac{f_s}{M_0} -i \frac{\Gamma}{q_o^2} - i \gamma q^2\right]\, .
\end{equation}
Similarly to the superfluid case, the dispersion relation of the pseudo-Goldstones is determined in terms of the pinning parameter and a single new independent relaxation rate.

\end{document}